\documentclass[prl,twocolumn,amsmath,amssymb,showpacs]{revtex4}
\usepackage{graphicx,epstopdf}
\usepackage{braket}

\newcommand{\Fkt}[1]{\,\mathsf {#1}}
\ifx\Tr\renewcommand{\Tr}{\Fkt{Tr}}
\else\newcommand{\Tr}{\Fkt{Tr}}
\fi

\begin{document}

\title{Optimal strategies for estimating the average fidelity of quantum
  gates}

\author{Daniel M. Reich}
\affiliation{Theoretische Physik, Universit\"{a}t Kassel,
  Heinrich-Plett-Str. 40, D-34132 Kassel, Germany} 
\author{Giulia Gualdi} 
\altaffiliation[present address: ]{QSTAR - {\it Quantum Science and
    Technology in Arcetri}, Largo Enrico Fermi 2, I-50125 Florence,
  Italy}  
\affiliation{Theoretische Physik, Universit\"{a}t Kassel,
  Heinrich-Plett-Str. 40, D-34132 Kassel, Germany}
\author{Christiane P. Koch}
\affiliation{Theoretische Physik, Universit\"{a}t Kassel,
  Heinrich-Plett-Str. 40, D-34132 Kassel, Germany} 
\email{christiane.koch@uni-kassel.de}
\date{\today}

\begin{abstract}
  We show that the minimum experimental effort to estimate the average
  error of a quantum gate scales as $2^n$ for $n$
  qubits 
  and requires classical computational resources $\sim n^22^{3n}$
  when no specific assumptions on the gate can be made.
  This represents a reduction by $2^n$ compared to the best
  currently available protocol, Monte Carlo characterization.
  The reduction comes at the price
  of either having to prepare entangled input states or obtaining
  bounds rather than the average fidelity itself. 
  It is achieved by applying Monte Carlo sampling to so-called
  two-designs or two classical fidelities. 
  For the specific case of Clifford gates, the original version of
  Monte Carlo characterization based on the channel-state isomorphism
  remains an optimal choice. We provide a classification of the 
  available efficient strategies to determine the average gate error 
  in terms of the number of required 
  experimental settings, average  number of actual measurements
  and classical computational resources.
\end{abstract}

\maketitle

\paragraph{Introduction}
The development of quantum technologies is currently facing a number
of obstacles. One of them is the difficulty to assess efficiently how
well a quantum 
device implements a desired operation. The corresponding performance
measure is the average fidelity or the average gate error which  can be
determined via  quantum process tomography~\cite{NielsenChuang}. 
Full process tomography scales, however, strongly exponentially in the
number of qubits and provides the full process matrix, i.e.,  
much more information than just the gate error. For
practical applications, a more targeted and less resource-intensive
approach  is required. 
Recent attempts at reducing the resources employ stochastic
sampling~\cite{FlammiaPRL11,daSilvaPRL11,ShabaniPRL11,SchmiegelowPRL11,MagesanPRL11,MagesanPRA12}.  
The process matrix can be estimated efficiently if it is
sparse in a convenient 
basis~\cite{MohseniPRA09,CramerNatComm10,ShabaniPRL11,SchmiegelowPRL11}. 
Randomized benchmarking, utilizing so-called unitary
$t$-designs~\cite{DankertPRA09}, is scalable 
when estimating the gate error for 
Clifford gates~\cite{MagesanPRL11,MagesanPRA12}. For
\textit{general} unitary operations, Monte Carlo sampling combined with the
channel-state isomorphism currently appears to be the most 
efficient approach~\cite{FlammiaPRL11,daSilvaPRL11}. It comes with the
advantage of separable input states. 
A second promising approach for general
unitaries utilizes state two-designs~\cite{BenderskyPRL08}.
They are given, for example,  by the states of $d+1$ mutually unbiased
bases~\cite{WoottersAnnPhys89}. 
Since only three out of the $d+1$ mutually unbiased bases consist of
separable states~\cite{LawrencePRA11}, 
entangled input states need to be prepared. 
The approaches of
Refs.~\cite{FlammiaPRL11,daSilvaPRL11} and~\cite{BenderskyPRL08} 
yield the average fidelity for general unitaries with an arbitrary, 
prespecified accuracy. They have been tested experimentally, albeit so 
far only for two- and three-qubit operations, 
without taking advantage of the protocols'
efficiency~\cite{SteffenPRL12,SchmiegelowPRL10}.  
Alternatively to estimating the average fidelity directly, an upper
and a lower bound can be obtained from two classical fidelities that
are evaluated  for
the states of two mutually unbiased bases~\cite{HofmannPRL05}. 
This approach has been employed in an experiment demonstrating quantum
simulation with up to 6 qubits~\cite{LanyonSci11}. 
When designing a quantum device, one is thus faced with a number of
options to determine the average gate errors, the optimality of which will
depend on specific experimental constraints and the required accuracy. 

Here, we provide a unified classification of the
currently available approaches in terms of the number and type of input
states and measurements that need to be available, the number of
actual experiments that need to be carried out and the classical
computational resources that are required.
We show that applying Monte Carlo estimation to the two-design 
protocol and the classical fidelities yields a reduction by a factor $2^n$
in resources for general unitary operations.
For the specific task of characterizing Clifford gates, the two
strategies are as efficient as Monte Carlo sampling combined with the
channel-state isomorphism~\cite{FlammiaPRL11,daSilvaPRL11} or
randomized benchmarking~\cite{MagesanPRL11,MagesanPRA12}.
The reduction in resources that we report here for general
unitary operations is made possible by avoiding 
the channel-state isomorphism: Instead of estimating 
the average fidelity in terms of a single state fidelity in Liouville
space, it is determined by a sum over state fidelities in Hilbert space. 
We have recently shown that a minimal set of states in Hilbert space
is sufficient for device characterization~\cite{ReichPRA13}. Therefore
the number of states that enter the sum is
determined only by the desired bounds.

\paragraph{Monte Carlo sampling} 
We first review Monte Carlo
estimation of the average fidelity 
as introduced in Refs.~\cite{FlammiaPRL11,daSilvaPRL11}
before applying it to 
two-designs~\cite{BenderskyPRL08,SchmiegelowPRL11} and  
classical fidelities~\cite{HofmannPRL05}. 
The fidelity of a quantum state or process is of the form  
$F=\Tr[\rho^{id}\rho^{act}]$ with $\rho^{id/act}$ the ideal/actual
state. Monte Carlo sampling estimates the quantity $F$ 
from a small random sample of measurements. $F$ thus needs to be
expressed in terms of measurement results. To this end, 
$\rho^{id}$ and $\rho^{act}$ are expanded in an orthonormal basis of 
Hermitian operators, yielding 
$F\sim \sum_k \Tr[\rho^{id}W_k]\Tr[\rho^{act}W_k]$. 
The measurement results are treated as a random variable $X$ taking
values $X_{\kappa}$ which occur with probability $\Fkt{Pr}(\kappa)$
($\sum_{\kappa=1}^{T}\Fkt{Pr}(\kappa) =1$ with $T$ the size of the
event space), i.e.,  
\begin{equation}
  \label{eq:MC}
  F= \sum_{\kappa=1}^{T} \Fkt{Pr}(\kappa) X_\kappa\,.
\end{equation}
Introducing 
$\chi_{\rho}(\kappa)=\Tr[\rho W_\kappa]$, 
$X_{\kappa}$ is given by
\[
X_\kappa=\frac{\chi_{\rho^{act}}(\kappa)}{\chi_{\rho^{id}}(\kappa)}
= \frac{\Tr[\rho^{act} W_\kappa]}{\Tr[\rho^{id}  W_\kappa]}
\quad\mathrm{and}\quad
\Fkt{Pr}(\kappa)=\frac{\chi_{\rho^{id}}(\kappa)^2}{\mathcal{N}}
\]
with $\mathcal{N}$ ensuring proper normalization.
$\Fkt{Pr}(\kappa)$ is also called relevance distribution. 
Two levels of stochastic sampling are involved in the 
Monte Carlo estimation of a fidelity:
According to Eq.~\eqref{eq:MC}, $F$ is
the expectation value of the random variable $X$ taking values
$X_\kappa$ with known probability $\Fkt{Pr}(\kappa)$.
However, the $X_\kappa$ cannot be accessed directly,
since they depend on another
random variable, the expectation value of $W_\kappa$ for
$\rho^{act}$. 
Due to the statistical nature of quantum measurements as well as
random errors in the experiment, it will be necessary to repeatedly measure
$W_\kappa$ in order to determine
$X_\kappa$. Assuming that the $X_\kappa$ have been determined with
sufficient accuracy, Monte Carlo sampling estimates their
expectation value $F$ from  a finite
number of realizations $L$, 
\begin{equation}
  \label{eq:MCapprox}
  F=\lim_{L\to\infty} F_{L} \quad\mathrm{with}\quad
  F_{L} = \frac{1}{L}\sum_{l=1}^L X_{\kappa_l}
\end{equation}
and $\kappa_L$ taking values between 1 and $T$.
$L$ is chosen such 
that the probability for $F_{L}$ to differ from 
$F$ by more than $\epsilon$ is less than $\delta$. 
The key point of the Monte 
Carlo approach is that $L$ depends only on the desired accuracy 
$\epsilon$ and confidence level $\delta$ and is
independent of the system size. Note that $\epsilon$ is lower
bounded by measurement and state preparation errors. Since, 
as quantum mechanical expectation values, the $X_{\kappa_l}$ are
known only approximately, also $F$ can be obtained only approximately:
$\tilde F_{L} = \frac{1}{L}\sum_{\kappa_l=1}^L \tilde X_{\kappa_l}$
(with $\tilde{X}_{\kappa_l}$ denoting the approximate values of
$X_{\kappa_l}$). Therefore, 
in addition to ensuring that  $F_{L}$ approximates $F$ with
a statistical error of at most $\epsilon$, one also has to
guarantee that $\tilde F_{L}$ approximates $F_{L}$ 
with the desired accuracy. This implies repeated measurements for a
given element $l$ ($l=1,\ldots,L$) of the Monte Carlo sample.
Denoting the number of respective measurements by $N_l$, the total
number of experiments is given by $N_{exp}=\sum_{l=1}^L N_l$.
It can be shown that a proper choice of $\langle N_l\rangle$ and
$\langle N_{exp}\rangle$ guarantees the approximations of $F_{L}$ by 
$\tilde F_{L}$ and of $F$ by $F_{L}$ to hold with the desired
confidence level. While $L$ is indepenent of system size, the choices
of  $\langle N_l\rangle$ and $\langle N_{exp}\rangle$ in general
depend on it.

For a quantum process, the average fidelity can be obtained by
Monte Carlo estimation when combining it with the channel-state
isomorphism~\cite{FlammiaPRL11,daSilvaPRL11}. 
$F_{av}$ is then expressed in terms of the entanglement fidelity $F_e$ 
via $F_{av}=(dF_e+1)/(d+1)$~\cite{HorodeckiPRA99}.
Since $F_e$ is a state fidelity in Liouville space and Liouville space
vectors correspond to Hilbert space operators,
this implies evaluation of $F_{av}$ with respect to an operator basis,
comparing input to output operators. Since a complete operator basis
consists of $2^{2n}$ elements and the size of the event space is given
by all possible combinations of input and output operators,
$T=2^{4n}$. The fact that only states, not operators can be
prepared as input is remedied by randomly selecting eigenstates of the
input operators. There are 6 eigenstates for the 3 Pauli
operators for each qubit. Therefore 
the number of  experimental settings, i.e., pairs of 
input state/output measurement operator,
is given by $N_{setting}=N_{input}\times
N_{meas}=6^n\cdot 2^{2n}$. The random selection of 
experimental settings requires classical 
computational resources $\mathcal C_{class}$ that scale as 
$n^2 2^{4n}$~\cite{FlammiaPRL11,daSilvaPRL11}. 
Although only some of the settings will be
selected, the ability to implement all of them in the experiment is
implied.  
Due to the statistical nature of measurements, all in all
$\langle N_{exp}\rangle$ runs of the experiment have to be carried
out. For the experimental implementation, $N_{setting}$, $\langle
N_{exp}\rangle$ and $\mathcal C_{class}$ thus characterize the procedure. 

\paragraph{Monte Carlo estimation of classical fidelities and
  two-designs} 
State fidelities in Hilbert space as opposed to a state fidelity
in Liouville space are sufficient to estimate the average fidelity of
an arbitrary quantum gate~\cite{HofmannPRL05,ReichPRA13}.
We therefore distinguish in a Monte Carlo event $\kappa_l$ between 
input states and measurement operators, $\kappa_l=(i_l,k_l)$. This
allows for applying Monte Carlo sampling to the classical fidelities
of Ref.~\cite{HofmannPRL05} and the two-design
approach~\cite{BenderskyPRL08}. 

The two classical fidelities
which yield an upper and a lower bound to the
average fidelity~\cite{HofmannPRL05} can be written as~\cite{ReichPRA13}
\begin{eqnarray}
  \label{eq:Fj}
  F_j &=& \frac{1}{d}\sum_{i=1}^d \nonumber
  \Tr[\rho_i^{j,id}\rho_i^{j,act}] \\ &=&
  \frac{1}{d}\sum_{i=1}^d 
  \Tr[U|\Psi^j_i\rangle\langle\Psi^j_i|U^+
  \mathcal{D}(|\Psi^j_i\rangle\langle\Psi^j_i|)] 
\end{eqnarray}
with $|\Psi^{j}_i\rangle$ the states of two mutually unbiased 
bases in $d$-dimensional Hilbert space ($i=1,\ldots,d,j=1,2,d=2^n$), 
$\mathcal D$ the dynamical map describing the actual
evolution, and $U$ the desired unitary.
Expanding the states $\rho_i^{j,id}$, 
$\rho_i^{j,act}$ in terms of Pauli operators,
Eq.~\eqref{eq:Fj} becomes 
\begin{equation}
  \label{eq:FjMC}
    F_j = \sum_{i=1}^d \sum_{k=1}^{d^2}\Fkt{Pr}^j(i,k)
    \frac{\chi^j_\mathcal{D}(i,k)}{\chi^j_U(i,k)}
\end{equation}
with characteristic function
$\chi^j_{U}(i,k)=\Tr[W_kU|\Psi^j_i\rangle\langle\Psi^j_i|U^+]$
and relevance distribution
$\Fkt{Pr}^j(i,k)=\frac{1}{d^2}[\chi^j_{U}(i,k)]^2$.
Note that $\Tr[W_kW_{k'}]=d\delta_{k,k'}$.
We show in the supplementary material that $\Fkt{Pr}^j(i,k)$ is properly
normalized such that we can estimate the two classical fidelities
$F_j$, and thus an upper and a lower bound to $F_{av}$, by Monte Carlo
sampling.  

The expression for the average fidelity when using two-designs,
given in terms of $d+1$ mutually unbiased
bases~\cite{BenderskyPRL08},  
\begin{equation}
  \label{eq:Ftwodesign}
  F^{2des}_{av}=\frac{1}{d(d+1)}\sum_{i=1}^{d(d+1)}
  \Tr[U|\Psi_{i}\rangle\langle\Psi_{i}|U^{+}
  \mathcal{D}(|\Psi_{i}\rangle\langle\Psi_{i}|)],
\end{equation}
is formally similar to Eq.~\eqref{eq:Fj}, i.e., it 
can be interpreted as the sum over $d+1$ classical 
fidelities.
Equation~\eqref{eq:Ftwodesign} can thus be rewritten
\begin{equation}
  \label{eq:FtwodesignMC}
  F^{2des}_{av}=\sum_{i=1}^{d(d+1)}\sum_{k=1}^{d^2}
  \Fkt{Pr}^{2des}(i,k)
  \frac{\chi^{2des}_\mathcal{D}(i,k)}{\chi^{2des}_U(i,k)}
\end{equation}
with characteristic function $\chi^{2des}_U$
analogous to $\chi_U^j$, and the   
relevance distribution differing only in normalization,
$\Fkt{Pr}^{2des}(i,k)=\frac{1}{d^2(d+1)}\left[\chi^{2des}_{U}(i,k)\right]^2$.
We show in the supplementary material that also $\Fkt{Pr}^{2des}(i,k)$
is properly normalized such that $F_{av}^{2des}$ can be estimated by
Monte Carlo sampling. 

\paragraph{Resources for estimating the gate error of
  general unitaries} 

\begin{table}[tb]
  \centering
  \begin{tabular}{|c|c|c|c|c|}
    \hline
    approach & $\mathcal C_{class}$&$N_{input}$ & $N_{setting}$ & $\langle N_{exp}\rangle$
    \\
    \hline
    A&$\mathcal{O}(n^22^{4n})$&
    $6^n$& 
    $\mathcal O(6^n2^{2n})$ & $\mathcal O(2^{2n})$ \\
    B&$\mathcal O(n^22^{4n})$& $2^n(2^n+1)$ & $\mathcal O(2^{4n})$& 
    $\mathcal O(2^n)$\\ \hline
    C & $\mathcal O(n^22^{3n})$&$2\cdot 2^n$  & 
    $\mathcal O(2^{3n})$& $\mathcal O(2^n)$\\
    \hline
  \end{tabular}
  \caption{Resources required for determining the average gate error
    of a general unitary 
    operation in terms of classical computational effort 
    $\mathcal C_{class}$ required for the random selection,
    number $N_{input}$ of input states that
    need to be prepared, the number of experimental settings
    $N_{setting}$ from which the actual experiments will be randomly
    chosen, and the average number $\langle N_{exp}\rangle$ of
    experiments to be performed. $N_{setting} = N_{input}\times
    N_{meas}$ with the number of measurement operators $N_{meas}=2^{2n}$
    for all cases (A: Monte Carlo sampling based on the channel-state
    isomorphism~\cite{FlammiaPRL11,daSilvaPRL11}; 
    B: Monte Carlo sampling for two-designs; 
    C: Monte Carlo sampling for classical fidelities).
  } 
  \label{tab:res_gen}
  \end{table}
Evaluating Eqs.~\eqref{eq:FjMC} or \eqref{eq:FtwodesignMC} by 
Monte Carlo estimation involves randomly
selecting $L$ times a pair $(i_l,k_l)$ of input state/measurement
operator. 
Compared to Refs.~\cite{FlammiaPRL11,daSilvaPRL11}, 
the number of input states is significantly reduced for the
two approaches based on state fidelities in Hilbert space. This
yields a correspondingly smaller number of settings that an 
experimentalist needs to be able to implement,
cf. Table~\ref{tab:res_gen}. Moreover, the smaller number of input
states reduces the classical computational
resources required for the random selection by a factor $2^n$
for the classical fidelities. This is due to 
$\mathcal C_{class}=N_{input}\times \mathcal C_{single}$ with 
$\mathcal C_{single}$ the classical computational cost for sampling 
a single state fidelity in Hilbert space ($\mathcal C_{single}\sim
n^22^{2n}$~\cite{daSilvaPRL11}). It reflects
the fact that the relevance distribution for the
classical fidelities depends on $\mathcal{O}(d^3)$ parameters 
whereas the relevance distribution of
Refs.~\cite{FlammiaPRL11,daSilvaPRL11} depends on 
$\mathcal{O}(d^4)$ parameters. 
The reduced number of parameters is 
sufficient to determine whether the actual 
evolution matches the desired unitary~\cite{ReichPRA13}. 

Analogously to Refs.~\cite{FlammiaPRL11,daSilvaPRL11}, we determine
the sample size $L$ by Chebychev's inequality. It provides an upper
bound for the probability of a random variable $Z$ with variance
$\sigma_Z$ to deviate from its mean, 
\begin{equation}
  \label{eq:Cheby}
  \Fkt{Pr}\left[\left|Z-\langle Z\rangle\right|
    \geq \sigma_Z/\sqrt{\delta}\right]\leq\delta
\end{equation}
with $\delta>0$. In our case, $\langle Z\rangle=F$ with
$F=F_{av}^{2des}$ or $F_j$, $Z=F_{L}=1/L\sum_{l=1}^LX_{\kappa_l}$,  and
$X_{\kappa_l}\equiv X_l=\chi_{\mathcal  D}(i_l,k_l)/\chi_U(i_l,k_l)$,  
cf. Eqs.~\eqref{eq:FjMC}, \eqref{eq:FtwodesignMC}. 
We show in the supplementary material that the variance of $X_l$ 
is smaller than one, and thus $\Fkt{var}(F_{L})\le 1/L$. 
Then the choice 
  $L=1/(\epsilon^2\delta)$
guarantees that the probability for
the estimate $F_{L}$ to differ from $F$ by more than
$\epsilon$ is smaller than $\delta$. Specifying the experimental
inaccuracy and choosing the confidence level thus determines the
sample size.

In order to estimate the number of required experiments, we first
determine the number of experiments for one setting, $N_l$.
For each $l$, the observable $W_{k_l}$ has to be measured $N_l$ 
times to account for the statistical nature of the measurement. The
corresponding approximation to $X_l$ is given by 
\begin{equation}
  \label{eq:tildeX_l}
  \tilde X_l= \frac{1}{\chi_U(i_l,k_l)}\frac{1}{N_l}\sum_{j=1}^{N_l} w_{lj}
\end{equation}
with $w_{lj}$ the measurement result for the $j$th repetition of
experimental setting $l$, equal to either +1 or -1 for Pauli
operators.
Since $\tilde X_l$ is given as the sum of independent random variables
$w_{lj}$, $N_l$ can be determined using Hoeff\-ding's inequality. 
It provides an upper bound for the probability of 
a sum $S=\sum_{i=1}^{n}Y_{i}$ of 
independent variables $Y_i$ with  $a_{i}\leq Y_{i}\leq b_{i}$ to 
deviate from its expected value by more than $\epsilon$,
\begin{equation}
  \label{eq:Hoeffding}
  \Fkt{Pr}\left(\left|S-\left\langle S\right\rangle \right|
    \geq \epsilon\right)\leq2\exp\left(-\frac{2\epsilon^{2}}
    {\sum_{i=1}^{n}\left(b_{i}-a_{i}\right)^{2}}\right)
\end{equation}
$\forall \epsilon>0$. In our case,
$S=\tilde F_{L}=\frac{1}{L}\sum_{l=1}^L\tilde X_l$ and, using
Eq.~\eqref{eq:tildeX_l}, 
$\sum_{i=1}^{n}\left(b_{i}-a_{i}\right)^{2}=
\sum_{l=1}^{L} 4N_{l}\left[LN_l\chi_U(i_l,k_l)\right]^{-2}$.
Inserting this into Eq.~\eqref{eq:Hoeffding}, it is obvious that the
choice 
\begin{equation}
  \label{eq:N_l}
  N_l =
  \frac{2}{L\epsilon[\chi_U(i_l,k_l)]^2}\log\left(\frac{2}{\delta}\right)
  = N_l(i_l,k_l)
\end{equation}
ensures the right-hand side of Eq.~\eqref{eq:Hoeffding} to be 
$\le \delta$. 
The setting $l$ is chosen with probability $\Fkt{Pr}^{j/2des}(i_l,k_l)$. The
average number of times that this specific
experiment (with input state $i_l$ and measurement operator $W_{k_l}$) is 
carried out is therefore given by 
\begin{eqnarray}
  \label{eq:expN_l}
  \langle N_l\rangle &=& 
  \sum_{i_l=1}^{d}\sum_{k_l=1}^{d^{2}}\Fkt{Pr}^j\left(i_l,k_l\right)N_{l}(i_l,k_l)\\
  &=&\frac{1}{d^{2}}\sum_{i_l=1}^{d}\sum_{k_l=1}^{d^{2}}[\chi^j_U(i_l,k_l)]^{2}
  \frac{4}{[\chi^j_U(i_l,k_l)]^{2}L\epsilon^{2}}\log\left(\frac{2}{\delta}\right)
  \nonumber \\ \nonumber
  & \leq & 1+\frac{2d}{L\epsilon^{2}}\log\left(\frac{2}{\delta}\right)
\end{eqnarray}
for the two classical fidelities ($j=1,2$). The same
$\langle N_l\rangle$ is obtained for the two-designs
due to normalization of $\Fkt{Pr}^{2des}\left(i,k\right)$. 
The total number of experiments that need to be carried out is then
estimated by  
\begin{eqnarray}
  \label{eq:Nexp}
  \langle N_{exp}\rangle = \sum_{l=1}^L \langle N_l\rangle
  &\leq& \nonumber
  L\left[1+\frac{2d}{L\epsilon^{2}}\log\left(\frac{2}{\delta}\right)\right]\\
  &\leq&
  1+\frac{1}{\epsilon^{2}\delta}+\frac{2d}{\epsilon^{2}}
  \log\left(\frac{2}{\delta}\right)\,. 
\end{eqnarray}
This number is sufficient to account for both the sampling error due to
finite $L$ and statistical experimental errors in the measurement results. 
Notably, $\langle N_{exp}\rangle\sim 2^{n}$ only,
i.e., the average number of experiments to estimate $F_{av}$
scales like that required for characterizing a general
pure  quantum  state~\cite{daSilvaPRL11}.
This represents a reduction by a factor $2^n$ compared to
Refs.~\cite{FlammiaPRL11,daSilvaPRL11}, cf. Table~\ref{tab:res_gen}.
These savings come at the expense of (i) obtaining only bounds on the
average fidelity when using two classical fidelities $F_j$ or (ii)
the necessity to prepare entangled input states when using  
two-designs. The latter scales quadratically in
$n$~\cite{BenderskyPRL08}. Even factoring this additional cost in,
Monte Carlo estimation of the average fidelity for a general unitary
operation using two-designs is
significantly more efficient than that based on the
channel-state isomorphism~\cite{FlammiaPRL11,daSilvaPRL11}. 

\paragraph{Resources for Monte Carlo estimation of
  Clifford gates} 
\begin{table}[tb]
  \centering
  \begin{tabular}{|c|c|c|c|c|}
    \hline
    approach &$\mathcal C_{class}$& $N_{input}$ & $N_{setting}$ 
    & $\langle N_{exp}\rangle$ \\     \hline
    A&$\mathcal{O}(1)$ & $6^{n}$ &
    $\mathcal O(6^n2^{n})$ & $\mathcal O(1)$ \\
    B &$\mathcal O(1)$& $2^n(2^n+1)$ & $\mathcal O(2^{3n})$
    & $\mathcal O(1)$\\ \hline
    C&$\mathcal{O}(1)$ & $2\cdot 2^n$  
    & $\mathcal O(2^{2n})$& $\mathcal O(1)$\\
    \hline
  \end{tabular}
  \caption{Resources required for determining the average gate error
    of a Clifford
    gate ($N_{setting} = N_{input}\times 2^n$). Symbols as in
    Table~\ref{tab:res_gen}. 
  }  
  \label{tab:res_clifford}
\end{table}
The scaling of $\langle N_{exp}\rangle$
with the number of qubits changes dramatically for
Clifford gates~\cite{FlammiaPRL11,daSilvaPRL11}. 
This is due to the property of Clifford gates to map
eigenstates of  a $d$-dimensional
set of commuting Pauli operators into eigenstates from 
the same set. The mutually unbiased bases in
Eqs.~\eqref{eq:Fj},~\eqref{eq:Ftwodesign} can
be chosen to be such eigenstates~\cite{LawrencePRA02}.
Given a generic eigenstate $|\Psi_i\rangle$
of a commuting set $\mathcal{W}$ of Pauli
operators, the characteristic function
of a Clifford gate, $U_{Cl}$, becomes
\begin{eqnarray}
  \label{eq:chi_Clifford}
  \chi_{U_{Cl}}(i,k)&=&\Tr[W_kU_{Cl}|\Psi_i\rangle\langle\Psi_i|U^\dagger_{Cl}]\\
  &=&\Tr[W_k|\Psi_j\rangle\langle\Psi_j|]=\left\{\begin{array}{cc}
      \pm1& \mbox{if} \quad W_k\in\mathcal{W}\\
      0 & \mbox{otherwise}   \\
    \end{array}
  \right.\,.\nonumber
\end{eqnarray}
The relevance distribution for Clifford gates, 
$\Fkt{Pr(i,k)}\sim[\chi_{U_{Cl}}(i,k)]^2$, is thus zero for
many settings and uniform otherwise. Since settings with 
$\Fkt{Pr(i,k)}=0$ will never be selected, the sampling 
complexity becomes independent of system size. 
Calculating  $\langle N_l\rangle$ according to Eq.~\eqref{eq:expN_l} 
for a uniform relevance distribution, and accounting for the correct
normalizations of $\Fkt{Pr(i,k)}$, 
$\langle N_l\rangle$ is found to be independent of $d$,
$\langle N_l\rangle\leq 1+2\log(2/\delta)/(L\epsilon^2)$, for all
three approaches. Consequently, also $\langle N_{exp}\rangle$ does not
scale with system size, $\langle N_{exp}\rangle\leq
1+1/(\epsilon^2\delta)+2\log(2/\delta)/\epsilon^2$, 
cf. Table~\ref{tab:res_clifford}. For Clifford gates, the three
approaches require therefore a similar, size-independent number of
measurements. 
A difference is found, however, for the
number of possible experimental settings. For each input state $i$,
there are only $d$ (instead of $d^2$) measurement operators $W_k$ with
non-zero expectation value. This leads to $N_{setting} =
N_{input}\times 2^n$ for Clifford gates,
cf. Table~\ref{tab:res_clifford}. The larger $N_{setting}$
required for approaches A and B in Table~\ref{tab:res_clifford}
comes with a potentially higher accuracy of the estimate which is,
however, limited by the experimental error of state preparation and
measurement. 

\paragraph{Conclusions}
We find the number of measurements required to estimate the gate
error for a general unitary to scale as $2^n$ for $n$ qubits. Our 
reduction by a factor of $2^{n}$ compared to the best currently
available approach~\cite{FlammiaPRL11,daSilvaPRL11} comes at the
expense of either determining bounds from two classical fidelities
instead of the average fidelity itself or  
allowing for entangled input states. 
For the classical fidelities,
the number of experimental settings that one needs to be able to
prepare and the classical computational resources required for the
sampling are also reduced  by a factor $2^n$.
All three approaches are significantly more efficient than 
traditional process tomography requiring of the order $2^{4n}$ measurements
and a computational cost scaling as $4^{6n}$.   
For the special case of Clifford gates, we find the number of
experiments to be 
independent of system size, just as in Monte Carlo estimation based on
the channel-state isomorphism~\cite{FlammiaPRL11,daSilvaPRL11} and
randomized benchmarking~\cite{MagesanPRL11,MagesanPRA12}. 

We have shown earlier~\cite{ReichPRA13} that the minimum number of
pure input
states for device characterization is of the order $2^n$. This
corresponds to the number of states required by the classical
fidelities. Monte Carlo sampling of  two classical fidelities
therefore realizes a strategy of minimal resources. 
Our comprehensive classification should allow an experimentalist to
choose the most suitable procedure
to determine the average fidelity, defined in terms of the number of
experimental settings, from which a Monte Carlo procedure randomly
draws realizations, and the actual number of experiments to be carried
out. 
\begin{acknowledgments}
  We would like to thank Daniel Burgarth, Tommaso Calarco and David
  Licht for helpful comments. 
  Financial support from the EC
  through the FP7-People IEF Marie Curie action Grant
  No. PIEF-GA-2009-254174 is gratefully acknowledged.
\end{acknowledgments}


\clearpage
\setcounter{equation}{0}





We provide here detailed proofs of the claims made in the paper.   

\subsection*{The relevance distribution for a classical fidelity,
  $\Fkt{Pr}^j(i,k)$, Eq.~(5) of 
  the main paper, is normalized.} 

In order to prove normalization of the relevance distribution
$\Fkt{Pr}^j(i,k)$, Eq.~(5) of  the main paper, we first show that
\begin{equation}
  \label{eq:1}
  \sum_{k=1}^{d^{2}}\Braket{\varphi_{i}|W_{k}|\varphi_{j}}
  \Braket{\varphi_{n}|W_{k}|\varphi_{m}}  =  d\delta_{im}\delta_{jn}
\end{equation}
with $\left\{|\varphi_j\rangle\right\}$ the canonical basis. 
Note that for each pair $\Ket{\varphi_{i}},\Ket{\varphi_{j}}$,  
there are exactly $d$ operators $W_{k}$ with
$\Braket{\varphi_{i}|W_{k}|\varphi_{j}}\neq 0$. 
This is seen easily in the bit representation
($W_k=\omega_k^1\otimes\ldots\otimes \omega_k^N$). 
For the $m$th qubit, the scalar product vanishes if there is a bit 
flip between the two states at the $m$th qubit and 
$\omega^m_{k}=\openone_{2}$ or $\sigma_{z}$.
Analogously, the scalar product vanishes 
if the $m$th qubit has the same value between the
two states and $\omega^m_{k}=\sigma_{x}$ or $\sigma_{y}$. 
There are thus only two choices of $W_{k}$ for each qubit that lead to
a non-zero scalar product. Repeating the argument over all $n$ qubits
gives exactly $2^{n}=d$ possible operators $W_{k}$ for which
$\Braket{\varphi_{i}|W_{k}|\varphi_{j}}\neq0$.

Consider now
$\Braket{\varphi_{i}|W_{k}|\varphi_{j}}\Braket{\varphi_{n}|W_{k}|\varphi_{m}}$ 
for a certain $W_{k}$ 
with $\Ket{\varphi_{i}}\neq\Ket{\varphi_{m}}$ and $\Ket{\varphi_{j}}$, 
$\Ket{\varphi_{n}}$  fixed. Since $\Ket{\varphi_{i}}\neq\Ket{\varphi_{m}}$
there exists a qubit, $l$,  where the two states differ. We differentiate
two cases for this qubit: 
\begin{enumerate}
\item $\Ket{\varphi_{j}}$ and $\Ket{\varphi_{i}}$
  take the same value on the $l$th qubit. 
  Then $\omega^l_{k}$ must be $\openone_{2}$ or $\sigma_{z}$
  for
  $\Braket{\varphi_{i}|W_{k}|\varphi_{j}}\Braket{\varphi_{n}|W_{k}|\varphi_{m}}$
  not to vanish. However, there exists an
  operator $W_{k'}$ such that  
  the contribution of the two operators to the sum,
  Eq.~\eqref{eq:1}, 
  \[
  \Braket{\varphi_{i}|W_{k}|\varphi_{j}}\Braket{\varphi_{n}|W_{k}|\varphi_{m}}
  +\Braket{\varphi_{i}|W_{k'}|\varphi_{j}}\Braket{\varphi_{n}|W_{k'}|\varphi_{m}}\,,
  \]
  vanishes. This operator 
  is identical to $W_{k}$ except that $\omega_{k'}^l=\openone$ if
  $\omega_k^l=\sigma_z$ and vice versa. Then
  \begin{eqnarray*}
    \Bra{\varphi_{i}}W_{k'} =  -\Bra{\varphi_{i}}W_{k}\quad\mathrm{and}\quad
    W_{k'}\Ket{\varphi_{m}} = W_{k}\Ket{\varphi_{m}}
  \end{eqnarray*}
  with the minus sign due to $\sigma_z$ on the $l$th qubit for either
  $W_k$ or $W_{k'}$. 
\item Alternatively, $\Ket{\varphi_{j}}$ and $\Ket{\varphi_{i}}$ take 
  different values on the $l$th qubit.
  Then $\omega^l_{k}$ must be $\sigma_{x}$
  or $\sigma_{y}$ for 
  $\Braket{\varphi_{i}|W_{k}|\varphi_{j}}\Braket{\varphi_{n}|W_{k}|\varphi_{m}}$
  not to vanish. Again, there exists an operator $W_{k'}$ such that  
  the contribution of the two operators to the sum,
  Eq.~\eqref{eq:1}, vanishes. This operator
  is identical to $W_{k}$ except that $\omega_{k'}^l=\sigma_x$ if
  $\omega_k^l=\sigma_y$ and vice versa. If
  $\Ket{\varphi_i^l}=\Ket{1}$ (and thus $\Ket{\varphi_m^l}=\Ket{0}$), 
  \begin{eqnarray*}
    \Bra{\varphi_{i}}W_{k'} = -i\Bra{\varphi_{i}}W_{k}\quad\mathrm{and}\quad
    W_{k'}\Ket{\varphi_{m}} = -iW_{k}\Ket{\varphi_{m}}\,.
  \end{eqnarray*}
  Otherwise, if $\Ket{\varphi_i^l}=\Ket{0}$ (and thus
  $\Ket{\varphi_m^l}=\Ket{1}$),  
  \begin{eqnarray*}
    \Bra{\varphi_{i}}W_{k'} = i\Bra{\varphi_{i}}W_{k}\quad\mathrm{and}\quad
    W_{k'}\Ket{\varphi_{m}} = iW_{k}\Ket{\varphi_{m}}\,.
  \end{eqnarray*}
In both cases, the terms in the sum cancel.
\end{enumerate}
Consequently, for each $W_{k}$ and 
$\Ket{\varphi_{i}},\Ket{\varphi_{j}},\Ket{\varphi_{m}},\Ket{\varphi_{n}}$ 
with $\Ket{\varphi_{i}}\neq\Ket{\varphi_{j}}$ there exists a ``pair
operator'' $W_{k'}$ which cancels the contribution of $W_{k}$ to
Eq.~\eqref{eq:1} such that 
$\sum_{k=1}^{d^{2}}\Braket{\varphi_{i}|W_{k}|\varphi_{j}}\Braket{\varphi_{n}|W_{k}|\varphi_{m}}=0$ if $\Ket{\varphi_{i}}\neq\Ket{\varphi_{m}}$. Repeating the
argument for $\Ket{\varphi_{j}}\neq\Ket{\varphi_{n}}$ leads to
\begin{widetext}
\begin{eqnarray*}
  \sum_{k=1}^{d^{2}}\Braket{\varphi_{i}|W_{k}|\varphi_{j}}
  \Braket{\varphi_{n}|W_{k}|\varphi_{m}} & = & 
  \delta_{im}\delta_{jn}\sum_{k=1}^{d^{2}}\Braket{\varphi_{i}|W_{k}|\varphi_{j}}
  \Braket{\varphi_{n}|W_{k}|\varphi_{m}} \\
  & = & \delta_{im}\delta_{jn}\sum_{k=1}^{d^{2}}\left|
    \Braket{\varphi_{i}|W_{k}|\varphi_{j}}\right|^{2}
\end{eqnarray*}
\end{widetext}
using Hermiticity of $W_{k}$ in the last step. Now validity of
Eq.~\eqref{eq:1} follows simply from the fact that there exist,
for each pair $\Ket{\varphi_{i}},\Ket{\varphi_{j}}$, exactly $d$
operators $W_{k}$ with non-vanishing
$\Braket{\varphi_{i}|W_{k}|\varphi_{j}}$, and, in the canonical basis,
these matrix elements are equal to one.

Expanding a general vector $\Ket{\Phi}$ in the canonical basis and 
using Eq.~\eqref{eq:1}, we find our second intermediate result, 
\begin{widetext}
\begin{eqnarray}
 \sum_{k=1}^{d^{2}}\left|\Braket{\Phi|W_{k}|\Phi}\right|^{2} &=&
  \sum_{k=1}^{d^{2}}\left|\sum_{i,j=1}^{d}c_{i}^{*}c_{j}
    \Braket{\varphi_{i}|W_{k}|\varphi_{j}}\right|^{2} 
  =  \nonumber
 \sum_{k=1}^{d^{2}}\sum_{i,j,n,m=1}^{d}c_{i}^{*}c_{j}c_{n}^{*}c_{m}\Braket{\varphi_{i}|W_{k}|\varphi_{j}}\Braket{\varphi_{n}|W_{k}|\varphi_{m}}\\
 &=& 
 \sum_{i,j,n,m=1}^{d}c_{i}^{*}c_{j}c_{n}^{*}c_{m}d\delta_{im}\delta_{jn}
 =  d\left(\sum_{i=1}^{d}\left|c_{i}\right|^{2}\right)=d\,.
  \label{eq:2}
\end{eqnarray}  
\end{widetext}

It is now straightfoward to prove normalization of
$\Fkt{Pr}^j(i,k)$, starting from the definition
\[
\Fkt{Pr}^j(i,k)=\frac{1}{d^2}\chi^j_{U}(i,k)^2
=\frac{1}{d^2}\left|\Tr[\rho_i^{j,id}W_k]\right|^2
\]
and 
$\rho_i^{j,id}=U\rho^j_iU^+=U|\Psi^j_i\rangle\langle\Psi^j_i|U^+$.
Then
\begin{widetext}
\begin{eqnarray}
\sum_{i=1}^{d}\sum_{k=1}^{d^{2}}\Fkt{Pr}^j\left(i,k\right) \nonumber
  &=& \frac{1}{d^{2}}\sum_{i=1}^{d}\sum_{k=1}^{d^{2}}
\bigg|\Tr\left[W_{k}U^+|\Psi^j_{i}\rangle\langle\Psi^j_{i}|U\right]\bigg|^2
\\
 & = & \frac{1}{d^{2}}\sum_{i=1}^{d}\sum_{k=1}^{d^{2}}\sum_{n,m=1}^{d}\nonumber
 \langle\Psi_{n}|UW_{k}U^+|\Psi^j_{i}\rangle\langle\Psi^j_{i}|\Psi_{n}\rangle
 \langle\Psi_{m}|\Psi^j_{i}\rangle\langle\Psi^j_{i}|U^{+}W_{k}U|\Psi_{m}\rangle\\
 & = & 
 \frac{1}{d^{2}}\sum_{i=1}^{d}\sum_{k=1}^{d^{2}}
 \langle\Psi^j_{i}|UW_{k}U^+|\Psi^j_{i}\rangle
 \langle\Psi^j_{i}|U^+W_{k}U|\Psi^j_{i}\rangle
=\frac{1}{d^{2}}\sum_{i=1}^{d}d=1\,,
\label{eq:normj}
\end{eqnarray}
\end{widetext}
where we have used Eq.~\eqref{eq:2} in the last line with
$|\Phi\rangle=U|\Psi^j_i\rangle$. 

\subsection*{The relevance distribution for two-designs,
  $\Fkt{Pr}^{2des}(i,k)$ is  normalized.}  

The normalisation of   $\Fkt{Pr}^{2des}(i,k)$
is verified analogously to the previous
subsection, 
\begin{widetext}
  \begin{eqnarray}    \label{eq:norm2des}
    \sum_{i=1}^{d\left(d+1\right)}\sum_{k=1}^{d^{2}}
    \frac{1}{d^{2}\left(d+1\right)}\Fkt{Pr}^{2des}\left(i,k\right) 
    & = &
    \frac{1}{d^{2}\left(d+1\right)}\sum_{i=1}^{d\left(d+1\right)}
    \sum_{k=1}^{d^{2}}\bigg|\Tr\left[W_{k} 
      U\Ket{\Psi_{i}}\Bra{\Psi_{i}}U^{+}\right]\bigg|^2\\ \nonumber
    & = &
    \frac{1}{d^{2}\left(d+1\right)}\sum_{i=1}^{d\left(d+1\right)}\sum_{k=1}^{d^{2}}
    \langle\Psi_{i}|U^{+}W_{k}U|\Psi_{i}\rangle
    \langle\Psi_{i}|UW_{k}U^{+}|\Psi_{i}\rangle
    =\frac{1}{d^{2}\left(d+1\right)}\sum_{i=1}^{d\left(d+1\right)}d=1\,,
  \end{eqnarray}
\end{widetext}
using again Eq.~\eqref{eq:2} with
$|\Phi\rangle=U|\Psi_i\rangle$. 

\subsection*{The variance of $X_l$ is smaller than one.}

$X_l$, the random variable of the top level of sampling in the Monte
Carlo estimation of the average fidelity, has 
been defined in the main paper as the ratio of the measurement
outcomes for the actual state and the ideal state, 
\[
X_l=\frac{\chi_{\mathcal D}(i_l,k_l)}{\chi_U(i_l,k_l)}\,.
\]  
We first show that the variance of each $X_{l}$ in the estimation of
the classical fidelities is not too large,
\begin{widetext}
\begin{eqnarray}
  \text{Var}\left(X_{l}\right) &=& \nonumber
  \mathbb{E}\left(X_{l}^{2}\right)-\mathbb{E}\left(X_{l}\right)^{2}
=\sum_{i=1}^{d}\sum_{k=1}^{d^{2}}\Fkt{Pr}^j\left(i,k\right)
  \left(\frac{\chi^j_{\mathcal  D}(i_l,k_l)}{\chi^j_U(i_l,k_l)}\right)^2
  -\left(\sum_{i=1}^{d}\sum_{k=1}^{d^{2}}\Fkt{Pr}^j\left(i,k\right)
    \frac{\chi^j_{\mathcal  D}(i_l,k_l)}{\chi^j_U(i_l,k_l)}\right)^{2}
  \\  & = & \nonumber
  \frac{1}{d^{2}}\sum_{i=1}^{d}\sum_{k=1}^{d^{2}}[\chi^j_{\mathcal  D}(i_l,k_l)]^{2}
  -\frac{1}{d^{2}}\left[\sum_{i=1}^{d}
    \Tr\left[U|\Psi^j_{i}\rangle\langle\Psi^j_{i}|U^{+}
      \mathcal{D}\left(|\Psi^j_{i}\rangle\langle\Psi^j_{i}|\right)\right]\right]^{2}\\
  & = & \frac{1}{d^{2}}\sum_{i=1}^{d}\sum_{k=1}^{d^{2}}
  \bigg|\Tr\left[W_{k}\mathcal{D}\left(|\Psi^j_{i}\rangle\langle\Psi^j_{i}|
    \right)\right]\bigg|^{2}-F_j^{2}
   \label{eq:varj}
\end{eqnarray}
\end{widetext}
with $F_j$ the classical fidelities ($j=1,2$) defined in Eq.~(5) of
the main paper. 
Since $0\leq F_{j}\leq1$, $0\leq F_{j}^{2}\leq1$.
The same is true for the first term. This can be seen as follows.
Each term $\mathcal{D}\left(|\Psi_{i}\rangle\langle\Psi_{i}|\right)$ can be
written as a density matrix $\rho_{i}$,
\begin{equation*}
  \mathcal{D}\left(|\Psi_{i}\rangle\langle\Psi_{i}|\right)=\rho_{i}=
  \sum_{n=1}^{d}\lambda_{n}^{\left(i\right)}
  |\phi_{n}^{\left(i\right)}\rangle\langle\phi_{n}^{\left(i\right)}|\,,
\end{equation*}
with eigenvectors 
$|\phi_{n}^{\left(i\right)}\rangle$ and 
eigenvalues $\lambda_{n}^{\left(i\right)}$.
Evaluating the trace for each $i$ in the corresponding eigenbasis yields
\begin{widetext}
\begin{eqnarray*}
  \frac{1}{d^{2}}\sum_{i=1}^{d}\sum_{k=1}^{d^{2}}
  \bigg|\Tr\left[W_{k}\mathcal{D}\left(|\Psi_{i}\rangle\langle\Psi_{i}|
    \right)\right]\bigg|^{2}
  & = & 
  \frac{1}{d^{2}}\sum_{i=1}^{d}\sum_{k=1}^{d^{2}}\left|\sum_{n,m=1}^{d}\langle
    \phi_{m}^{\left(i\right)}|W_{k}\lambda_{n}^{\left(i\right)}|\phi_{n}^{\left(i\right)}\rangle
    \langle\phi_{n}^{\left(i\right)}|\phi_{m}^{\left(i\right)}\rangle\right|^{2}\\
  & = & \frac{1}{d^{2}}\sum_{i=1}^{d}\sum_{k=1}^{d^{2}}\left|\sum_{n=1}^{d}\langle\phi_{n}^{\left(i\right)}|W_{k}\lambda_{n}^{\left(i\right)}|\phi_{n}^{\left(i\right)}\rangle\right|^{2}
 \leq 
 \frac{1}{d^{2}}\sum_{i=1}^{d}\sum_{n=1}^{d}\left(\lambda_{n}^{\left(i\right)}\right)^{2}\sum_{k=1}^{d^{2}}\left|\langle\phi_{n}^{\left(i\right)}|W_{k}|\phi_{n}^{\left(i\right)}\rangle\right|^{2}\,.
\end{eqnarray*}
Using Eq.~\eqref{eq:2} with $\Ket{\Phi}=|\phi_n^{(i)}\rangle$ and
$\Tr[\rho_i^2]=\sum_{n}^d\left(\lambda_n^{(i)}\right)^2\leq 1$, we obtain  
\begin{equation}
  \label{eq:sumj}
  \frac{1}{d^{2}}\sum_{i=1}^{d}\sum_{k=1}^{d^{2}}
  \left|\Tr\left[W_{k}\mathcal{D}\left(|\Psi_{i}\rangle\langle\Psi_{i}|
      \right)\right]\right|^{2}
  \leq \frac{1}{d}\sum_{i=1}^{d}\sum_{n=1}^{d}
  \left(\lambda_{n}^{\left(i\right)}\right)^{2}
  \leq\frac{1}{d}\sum_{i=1}^{d}1=1\,.  
\end{equation}
\end{widetext}
Hence $\text{Var}\left(X_{l}\right)$ is the difference between two
numbers in the interval $\left[0,1\right]$ and therefore smaller
than one.
Just as Eq.~\eqref{eq:norm2des} was derived by replacing in
Eq.~\eqref{eq:normj} the summation limit of $d$ by $d(d+1)$ in the sum
over states $i$ and utilizing the normalization of $\Fkt{Pr}^{2des}$, 
the proof of
$\text{Var}\left(X_{l}\right)\leq1$ for the two-designs proceeds
analogously to Eqs.~\eqref{eq:varj} and \eqref{eq:sumj}.


\end{document}